\documentclass[conference]{IEEEtran}
\IEEEoverridecommandlockouts
\usepackage{cite}
\usepackage[ruled,vlined]{algorithm2e}
\usepackage{amsmath,amssymb,amsfonts}
\usepackage{algpseudocode}
\usepackage{graphicx}
\usepackage{textcomp}
\usepackage{xcolor}
\usepackage{multirow}
\usepackage{makecell}
\usepackage{footnote}
\def\BibTeX{{\rm B\kern-.05em{\sc i\kern-.025em b}\kern-.08em
    T\kern-.1667em\lower.7ex\hbox{E}\kern-.125emX}}
\begin{document}

\title{Multiplier with Reduced Activities and Minimized Interconnect for Inner Product Arrays}

 \author{\IEEEauthorblockN{Muhammad Usman\IEEEauthorrefmark{1}, Jeong-A Lee\IEEEauthorrefmark{1} and
 	Milo\v{s} D. Ercegovac\IEEEauthorrefmark{2}}
 	
 	\IEEEauthorblockA{\\
 	\IEEEauthorrefmark{1}Department of Computer Engineering, Chosun University, Gwangju, Republic of Korea.\\
 		Email: {usman}@chosun.kr, jalee@chosun.ac.kr\\
		\IEEEauthorrefmark{2}Computer Science Department, University of California, Los Angeles, CA, USA.\\
 		Email: {milos}@cs.ucla.edu\\
 		}}
\maketitle

\begin{abstract}
We present a pipelined multiplier with reduced activities and minimized interconnect based on online digit-serial arithmetic. The working precision has been truncated such that $p<n$ bits are used to compute $n$ bits product, resulting in significant savings in area and power. The digit slices follow variable precision according to input, increasing upto $p$ and then decreases according to the error profile. Pipelining has been done to achieve high throughput and low latency which is desirable for compute intensive inner products. Synthesis results of the proposed designs have been presented and compared with the non-pipelined online multiplier, pipelined online multiplier with full working precision and conventional serial-parallel and array multipliers. For $8, 16, 24$ and $32$ bit precision, the proposed low power pipelined design show upto $38\%$ and $44\%$ reduction in power and area respectively compared to the pipelined online multiplier without working precision truncation.
\end{abstract}


\section{Introduction}\label{sec: Intro}

\begin{figure}[ht]
	\begin{center}
\includegraphics[viewport=60 10 1225 850,scale=0.18]{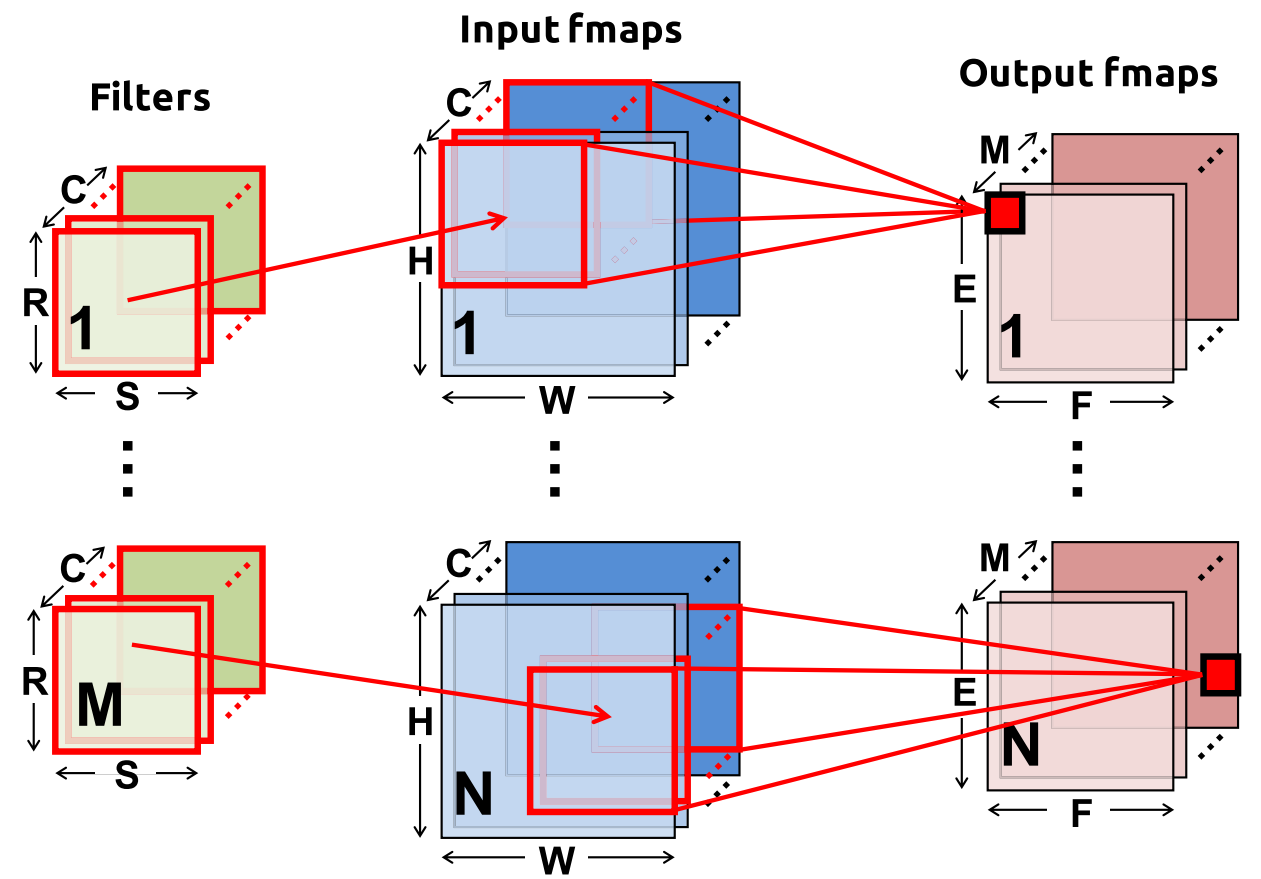}
	\end{center}
	\caption{Convolution neural network concept diagram. \cite{chen2016eyeriss}}
 	\label{fig: fmap}
\end{figure}

Convolutional neural network (CNN), which is a common form of DNN, provides superior accuracy in variety of applications including facial recognition, sentiment analysis, advertising, understanding the climate etc. \cite{cheng2019deep, rhanoui2019cnn, gao2018attention}. The block diagram of a convolutional neural network is depicted in Fig.~\ref{fig: fmap}. 

\begin{figure}[ht]
	\begin{center}
\includegraphics[viewport=75 10 1200 480,scale=0.2]{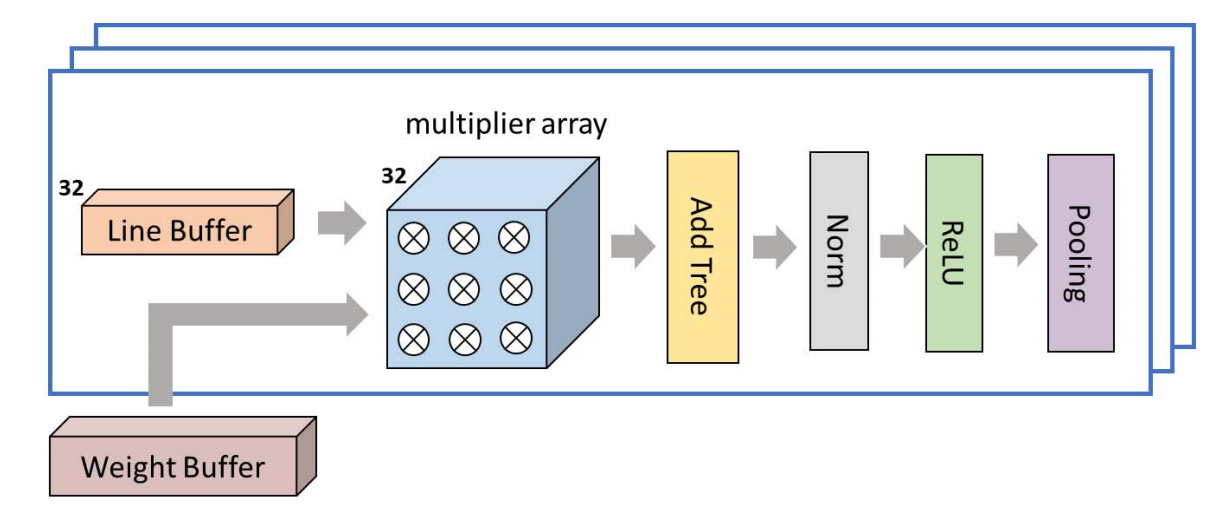}
	\end{center}
	\caption{Block diagram of matrix multiplication engine \cite{bai2018cnn}}
 	\label{fig: CNN_DWC}
\end{figure}

The classification accuracy of the CNNs are continuously improving and so is their complexity, therefore, the handheld devices with limited power resources find it cumbersome to employ massive inner product arrays of the CNNs \cite{khan2020survey}. Fig.~\ref{fig: fmap}, conceptually illustrates the computation of a CNN layer. 
To increase the evaluation throughput and efficiency, several CNN accelerators have also been proposed in the recent years. An accelerator based on depthwise separable convolution for FPGA has been proposed in \cite{bai2018cnn}. The convolution operation of all layers is performed in a computing unit named as matrix multiplication engine shown in Fig.~\ref{fig: CNN_DWC}. 

\begin{figure}[ht]
	\begin{center}
\includegraphics[viewport=5 10 390 130,scale=0.62]{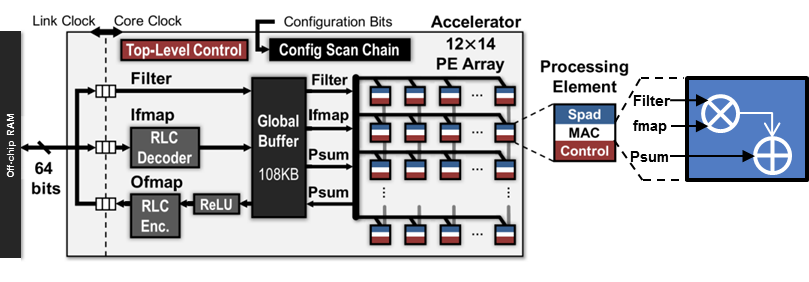}
	\end{center}
	\caption{Architecture of EYERISS \cite{chen2016eyeriss}}
 	\label{fig: DNN}
\end{figure}

Another well known accelerator for CNNs named EYERISS \cite{chen2016eyeriss}, consists of 2-D arrays of processing elements (PE) to compute inner product as shown in Fig.~\ref{fig: DNN}. 
In other accelerators including the ones mentioned, researchers provide an efficient architecture to compute millions of inner products. The product computation which is carried out using multipliers can thus be regarded as the key processing unit in these accelerators and is often regarded as the bottleneck \cite{venieris2018toolflows}. Multiplier, which is a complex hardware as compared to simpler operators such as adder, if implemented efficiently, can improve the performance of the inner product computation in both matrix multiplication engine and the PEs in EYERISS to a great extent.

To this end, we present a low-power pipelined multiplier for the inner product computation in compute intensive algorithms with reduced signal activities and minimized interconnect. The proposed multiplier is based on online arithmetic which computes in most significant digit first (MSDF) manner and has following properties \cite{ercegovac2004digital,ercegovac2017left,ercegovac2020}:
\begin{itemize}
    \item Each iteration, which is referred to as a \emph{step}, has been unrolled, which allows to instantiate only the required number of bit-slices in a certain cycle. Thus gradual bit-slice activation/deactivation has been exploited to reduce both static and leakage power.
    \item Unlike conventional online multiplier, in which $n$-bits must be implemented to obtain $n$-bit result, the proposed multiplier implements $p<n$-bits to compute $n$-bits result accurately.
    \item The proposed design can perform variable precision arithmetic by controlling the number of iterations, which is to simply stop the operation upon achieving the desired precision.
    \item The step time is kept short and consistent by computing and keeping the residual in redundant form.
    \item Due to nature of online operations, the successive online operations can be pipelined/ overlapped regardless of the data dependency, so that the latter operation does not have to wait for the completion of the former operation, as in the case of conventional arithmetic where overall latency is governed by the latencies of individual operations.
\end{itemize}

The remainder of paper is laid out as follows: A brief overview of online arithmetic and online multiplier has been presented in section \ref{sec: Online}. Details of the proposed implementation of low-power online multiplier has been discussed in section \ref{sec: proposed}. The results of the synthesis have been presented in section \ref{sec: Results}, and the paper is concluded in section \ref{sec: Conclusion}.

\section{Online Arithmetic and Online Multiplier} \label{sec: Online}
Algorithms are mostly implemented in digit-parallel form using conventional computation, resulting in an increased power and area footprint due to full-bandwidth interconnection data-paths. On the contrary, the interconnection bandwidth in the digit-serial computations is greatly reduced \cite{online_overview}. In conventional arithmetic, the succeeding operation can be started only after the completion of the preceding operation. This method of computation results in substantial resource wastage and an added delay because the successive layers in CNN or signal processing applications rely on the results of previous layer and operator respectively. 
 
These issues are circumvented in \emph{online arithmetic} \cite{ercegovac2004digital, online_overview}. It takes the input digits and produces the result, digit-by-digit manner starting from most significant digit first, which decreases the effect of data dependencies on the latency and throughput  \cite{ercegovac2017left}. The computation of the output is performed using the redundant representation of the partial results and is independent of operand's length. Redundant number systems are used in the evaluation of online arithmetic algorithms which adds flexibility in selection of the output because a given value can have several representations \cite{ercegovac2004digital}. Commonly used redundant number system is the signed digit (SD) which is a weighted radix-$r$ number system using which at any given iteration $j$, the fixed point digit  $x_j$ is represented by two single bits, $x_j^+$ and $x_j^-$, and $x_j = x_j^+ - x_j^-$. The numerical value of the digit $x$ is denoted as $x[j]$. The corresponding online form of the input operands and the product digit $z$ at iteration $j$ is represented as:

\begin{equation}
x[j] = \sum_{i=1}^{j+\delta}x_{i}r^{-i}\;\;\;\;\;\; \\
y[j] = \sum_{i=1}^{j+\delta}y_{i}r^{-i} \;\;\;\;\;\; \\
z[j] = \sum_{i=1}^{j}z_{i}r^{-i}
\end{equation}

\begin{figure}[ht]
	\begin{center}
\includegraphics[viewport=5 10 220 155,scale=1.0]{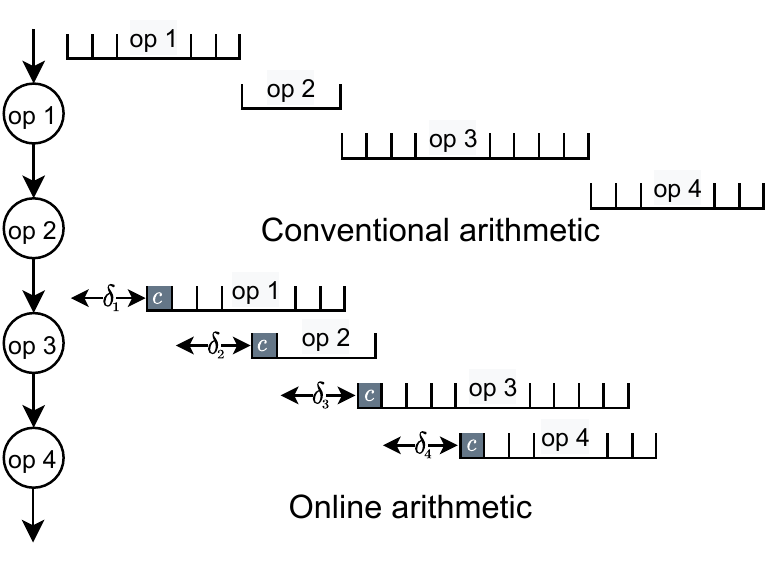}
	\end{center}
	\caption{Timing comparison of conventional and online arithmetic for sequence of operations assuming $\delta_i$ = $3$ and compute cycle $c$ = $1$. Conventional arithmetic has to wait for the completion of previous computation. Using online arithmetic, the successive operation can be started regardless of the data dependency as soon as $\delta$ digit results of the previous operation have been produced.}
 	\label{fig: Timing}
\end{figure}

The iteration is shown in the square bracket, while the subscripts show the digit index.
Online arithmetic algorithms confronts a fixed small delay, known as online delay ($\delta$), during which enough input digits  are accumulated to determine output within error bounds, after which the output is produced one digit per cycle. Computing most significant digit first allows arithmetic operations to be overlapped in order to achieve parallelism regardless of data dependency. The successive operation that requires the output of the previous operator can start computation as soon as few digits have been obtained and does not have to wait for the completion of entire result. The timing of conventional and online arithmetic for successive operations is shown in Fig.~\ref{fig: Timing}. 

\begin{figure}[ht]
	\begin{center}
\includegraphics[viewport=18 8 265 180,scale=1.0]{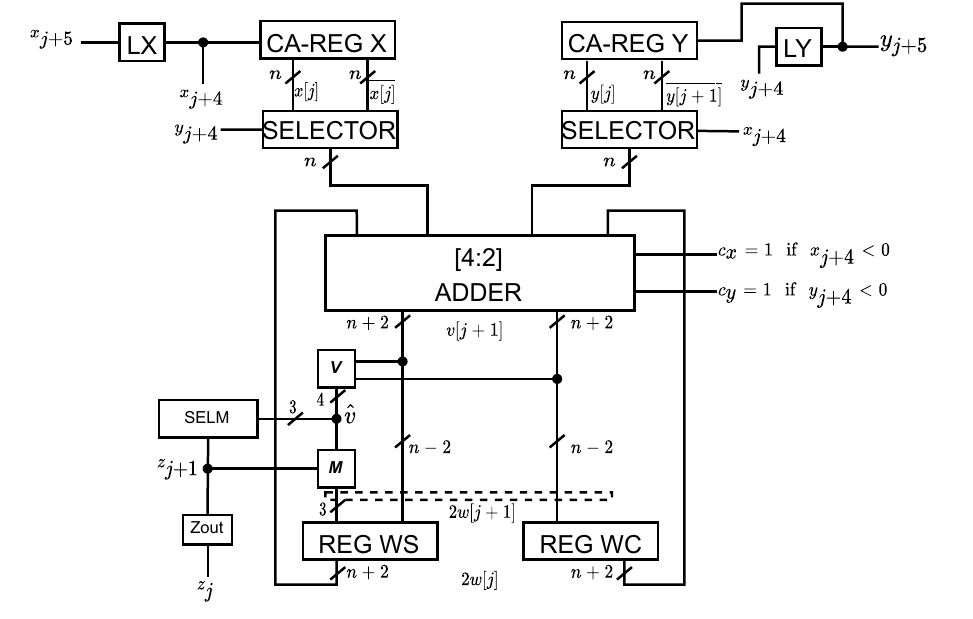}
	\end{center}
	\caption{Conventional radix-2 online multiplier \cite{ercegovac2004digital}. All $n$-bit slices remain active during all iterations.}
 	\label{fig: R2Mult}
\end{figure}

\begin{figure*}[ht]
	\begin{center}
\includegraphics[viewport=10 5 710 210,scale=0.70]{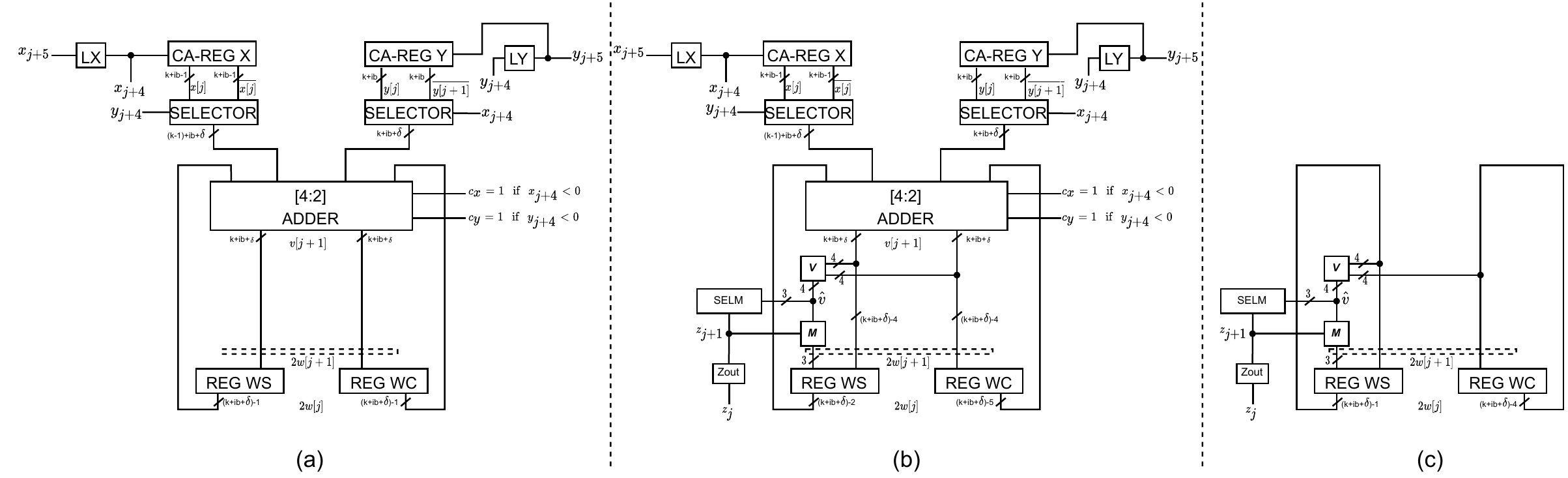}
	\end{center}
	\caption{(a) Initialization stage: inputs are received and number of bit-slices increase in each cycle. No output is produced, therefore, $V$, $M$, $SELM$ modules are not instantiated. (b) Recurrence stage: All modules are functional and follow gradual activation/ deactivation of bit-slices. (c) Last $\delta$ cycles: Input is $0$, therefore, the input reception modules are removed from the design. }
 	\label{fig: proposed}
\end{figure*}
The design of online multiplier (OLM) as shown in Fig.~\ref{fig: R2Mult}, has been presented in \cite{ercegovac2004digital}. The algorithm has an online delay $\delta=3$ and the number of fractional most significant digits (MSDs) used to select the output in the selection function are $t=2$. Like any other online algorithm, the output of the online multiplier is determined on the basis of partial input information which is composed in the residual. The scaled residual can be defined as:
\begin{equation} \label{eq: scaledresidual}
    w[j] = r^j (x[j]\cdot y[j]-z[j]),
\end{equation}
which can be deduced to obtain recurrence:
\begin{equation}
\begin{split}
w[j+1] & = rw[j] +(x[j] y_{j+1+\delta} + y[j+1]  x{j+1+\delta} )r^{-\delta} \\
& \;\;\;\;- q_{j+1},
\end{split}
\end{equation}
and is further decomposed into (\ref{eq: recurrence2}) and (\ref{eq: recurrence3}):
\begin{equation}\label{eq: recurrence2}
  v[j] = rw[j] +(x[j] y_{j+1+\delta}+ y[j+1] x_{j+1+\delta} )r^{-\delta} \\
\end{equation}
\begin{equation}\label{eq: recurrence3}
w[j+1] = v[j] - z_{j+1}
\end{equation}

For $r=2$ and $\delta = 3$ $v[j]$ in \eqref{eq: recurrence2} can be rewritten as:
\begin{equation}\label{eq: recurrence4}
v[j] = 2w[j] +(x[j] y_{j+4}+ y[j+1] x_{j+4} )2^{-3} \\
\end{equation}


The inputs and the residual are initialized as $x[-3]=y[-3]=w[-3]=0$. 
The algorithm executes for $n+\delta$ iterations which can be divided into three parts: (i) \emph{Initialization}: Its execution length is equal to $\delta$ during which one input digit is introduced and the output digit $z_j$ remains zero (therefore, the term $z_{j+1}$ is not present in \eqref{eq: recurrence3} during initialization). (ii) \emph{Recurrence}: It executes for $n-\delta$ cycles and produces an output digit which are selected by \emph{SELM} module using estimate of $v[j]$, denoted as $\widehat{v}[j]$, with $t=2$, as shown in Eq.~\eqref{eq: SELM}.
(iii) \emph{Last $\delta$ cycles}: Its execution length is also equal to $\delta$ and produces an output digit using the same procedures as in recurrence stage, however, the input digits are equal to zero. Details of modules used in each stage are provided in Section \ref{sec: proposed}.

\begin{equation}\label{eq: SELM}
    z_{j+1} =SELM(\widehat{v}[j])= \left\{\begin{matrix} \vspace{2mm} 
1 \;\;\;\;\; if \;\;\;\;\; \frac{1}{2} \leq \widehat{v} \leq \frac{7}{4}\\ \vspace{2mm}
0  \;\;\;\;\;if  \; -\frac{1}{2} \leq \widehat{v} \leq \frac{1}{4}\\
-1\;\;\;\;\: if \;\: -2 \leq \widehat{v} \leq -\frac{3}{4} 
\end{matrix}\right.
\end{equation}

\begin{figure}[!ht]
	\begin{center}
\includegraphics[viewport=32 10 550 990,scale=0.40]{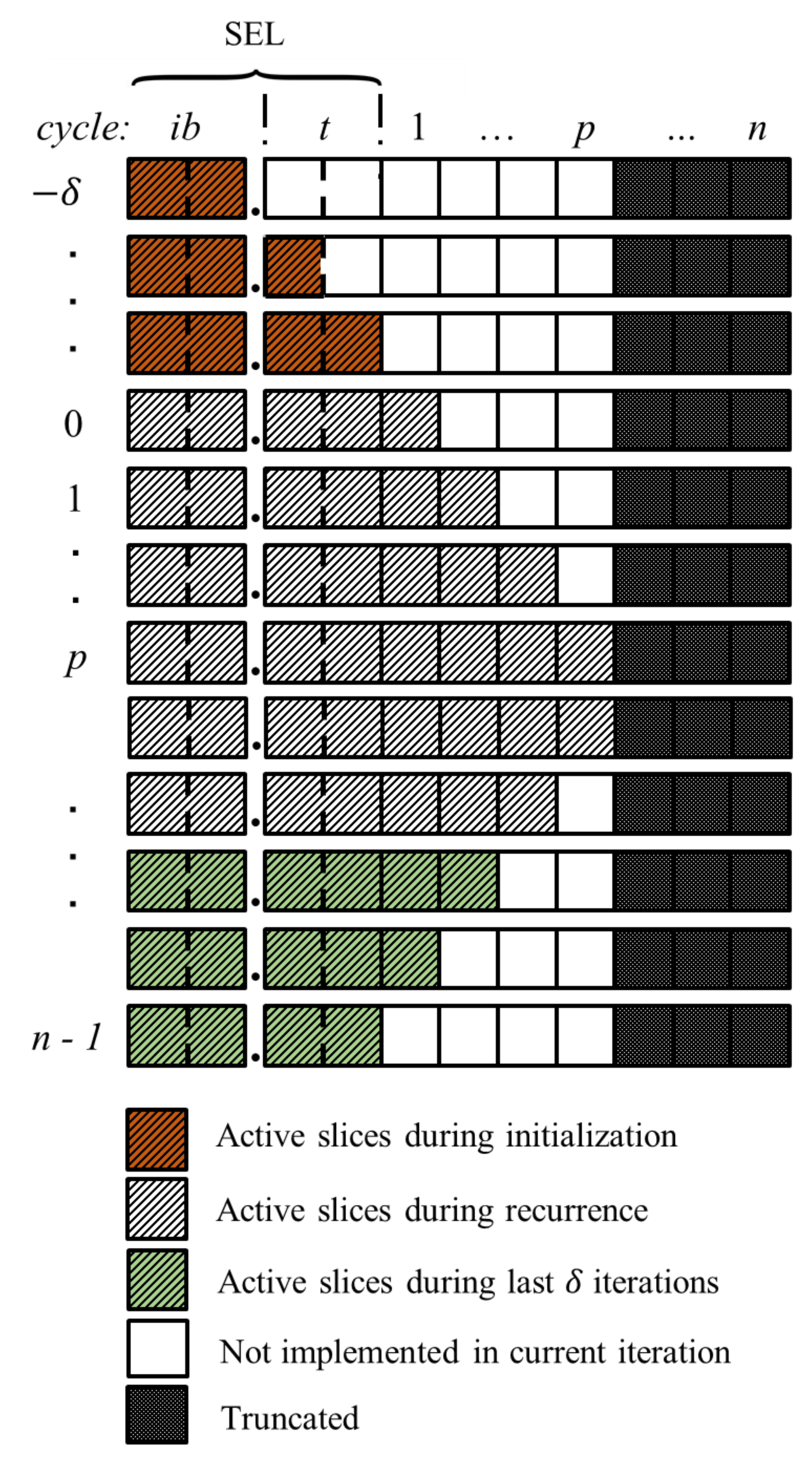}
	\end{center}
	\caption{Signal activity of radix-2 online multiplication algorithm with maximum truncated precision of \emph{p} with $\delta=3$, $2$ integer bits (\emph{ib}) and \emph{t} $=2$ }
 	\label{fig: OLactivity}
\end{figure}
\section{Reducing Signal Activities in the Conventional Online Multiplier}\label{sec: proposed}
In the proposed work, we pipeline the algorithm so as to compute multiple input vectors at the same time resulting in increased throughput. Furthermore, maximum working precision has been truncated and in efforts to reduce the signal activities of the conventional online multiplier, to realize the gradual activation/deactivation of digit-slices scheme has been utilized and only the required digit slices are implemented in each stage. The required number of active modules would be increased gradually according to precision of input operands from $1$ to $n$ due to online property. Furthermore, the number of digit slices does not necessarily need to be equal to the required precision, because after $p$ cycles, the new incoming digit affects more than necessary digits (`$t$') for the selection modules and hence bit slices after steps $p+1$ can be truncated. The truncation of bit slices introduces an error in the residual, which propagates to left due to the term $2w[j]$. The error propagation shows some pattern which is governed by the type of adder and radix $r$ used. The bit slices in the affected positions are not implemented in the forthcoming iterations. 
Minimum number of bit slices to have a valid selection function using an estimate of $t$ fractional bits for online multiplier having $[4:2]$ adder with radix $r=2$ is given by relation \ref{eq: pcalc1}, developed in \cite{ercegovac2004digital}.
\begin{equation}\label{eq: pcalc1}
    p = \left \lceil \frac{2n +\delta +t}{3}  \right \rceil
\end{equation}
The reduced working precision $p$ in relation \eqref{eq: pcalc1}, ensures accuracy of `$t$' fractional MSDs. All stages require similar basic components including registers, append units, redundant carry-save adders (CSA), converters from redundant to conventional representations (OTFC \cite{ercegovac1987fly}), ($ib+t$) width carry propagate adders (CPA) for the digit selection function (SELM) \cite{ercegovac2004digital,ercegovac2020}. 
The logic for each stage of the proposed implementation has following properties: (module names are referred from Fig.~\ref{fig: R2Mult}) 

\begin{itemize}
    \item \emph{Initialization stage}- Incoming digits are appended in respective \emph{CA-REG} register and no output is produced in this stage. Therefore, the units required for output generation including \emph{V}, \emph{M} and \emph{SELM} modules are not implemented. A block diagram of this stage has been presented in Fig.~\ref{fig: proposed} (a). 
    
    \item \emph{Recurrence}- The remaining input digits are received, multiplied using \emph{SELECTOR} ($4-$to$-1$ multiplexer) module and added to the residual by the [$4:2$] CSA \emph{ADDER}. Estimate of residual $\widehat{v}$ is calculated by the CPA module (\emph{V}), an output digit ($z_{j+1}$) is selected by the \emph{SELM} module and subtracted from $\widehat{v}$ by the \emph{M} block. The number of modules in the recurrence stage is similar to that of the conventional online multiplier as can be seen in Fig.~\ref{fig: proposed} (b). However, the number of bit slices are not $n$-bit wide and follows gradual activation/ deactivation pattern.
    
    \item In the \textit{last $\delta$ iterations}, all input operands are $0$, reducing relation \eqref{eq: recurrence2} to $2w[j]$, and output is produced in each cycle. Therefore, modules for appending input digits, \emph{SELECTOR} and \emph{ADDER} are not instantiated in this stage as shown in Fig.~\ref{fig: proposed} (c). 
\end{itemize}
These signal activity in these stages is highlighted in Fig.~\ref{fig: OLactivity}.

Additionally, the circuit for implementing the logic of the modules is carefully optimized and every non-functional register or adder in any stage has been removed. The conversion of signed input digits to the conventional number system is carried without additional delay by the OTFC modules during initialization and recurrence stages.

\section{Results}\label{sec: Results}
The implementation has been done for $n = 8, 16, 24$ and $32$ bits in pipelined manner to exploit the properties of online arithmetic. The synthesis results for the non-truncated pipelined and the proposed low power design have been compared in Table.~\ref{Comparison1}. The comparison of the proposed multiplier with conventional arithmetic based multiplier for $n=8$ has been presented in Table.~\ref{Comparison2}. The conventional multipliers can employ recoding techniques to reduce the number of partial products and use radix-$4$ implementation which results in a decreased latency. However, the cycle time of such implementation is increased. The synthesis was performed using the open synthesis suite Yosys 
from a verilog description. The area is reported relative to the area of a single NAND gate for the MCNC library assuming $20$ MHz clock and  $V_{dd} = 5$V from the following dictionary (BUFF 0.0, NOT 0.67, NAND 1.0, NOR 1.0, AND 1.33, OR 1.33, XOR 2.0, XNOR 1.66) calculated by \cite{vasicek2014evolutionary}. The combinational logic power, assuming a zero delay model was estimated using Berkley-SIS \cite{brayton2010abc}. Table.~\ref{Comparison3} shows the number of clock cycles required to multiply elements of vectors forming a stream. The conventional designs have to wait until the full precision result has been calculated, however, the proposed pipelined takes $n+\delta+1$ cycles to fill the pipeline and produce the output of first vector, after which an output is produced in each clock cycle. For large number of vectors i.e., $k>>n$, this delay is negligible and thus have significant advantage over the conventional multipliers.

\begin{table}[!ht]
\renewcommand{\arraystretch}{1.3}
\caption{Comparison of area and power estimates of $n$ bit pipelined designs for the radix-2 online multiplier with full and reduced working precision.}
\resizebox{0.5\textwidth}{!}{
\begin{tabular}{l|cccc|c}
\Xhline{3\arrayrulewidth}
\multirow{2}{*}{\textbf{Working Precision}} & \multicolumn{4}{c|}{\textbf{$n$}} & \multirow{2}{*}{\textbf{Resources}} \\ 
\cline{2-5}& \textbf{8} & \textbf{16} & \textbf{24} & \textbf{32} \\ \hline\hline
Full & 432 & 1734 & 2906 & 4844 & \multirow{3}{*}{{Latches}} \\ 
Reduced & \textbf{315} & \textbf{976} & \textbf{1906} & \textbf{3162}   \\ 
Savings (\%) & 27.08 & 31.93 & 34.41 & 34.72   \\ \hline

Full & 2385 & 1903 & 18402 & 30869 & \multirow{3}{*}{{Nodes}} \\ 
Reduced & \textbf{1786} & \textbf{5898} & \textbf{18455} & \textbf{17801}   \\ 
Savings (\%) & 25.11 & 34.51 & 37.87 & 40.21   \\ \hline

Full & 4474 & 16851 & 34617 & 58204 & \multirow{3}{*}{{Edges}} \\ 
Reduced & \textbf{3395} & \textbf{11363} & \textbf{22112} & \textbf{35759}   \\ 
Savings (\%) & 24.11 & 32.56 & 36.12 & 38.56   \\ \hline

Full & 2629.39 & 10529.32 & 21556.31 & 36217.59 & \multirow{3}{*}{{Area}} \\ 
Reduced & \textbf{1947.91} & \textbf{6432.94} & \textbf{12461.77} & \textbf{20133.69}   \\ 
Savings (\%) & 25.91 & 38.90 & 42.18 & 44.40   \\ \hline

Full & 25812.80 & 95179.70 & 194340.50 & 325686.80 & \multirow{3}{*}{{Power ($\mu$ watts)}} \\ 
{Reduced} & \textbf{18695.50} & \textbf{62720.40} & \textbf{122039.00} & \textbf{199687.70}   \\ 
{Savings (\%)} & 27.57 & 34.10 & 37.20 & 38.68  \\ \Xhline{3\arrayrulewidth}

\end{tabular}}
\label{Comparison1}
\vspace{-3mm}
\end{table}

\begin{table}[!ht]
\renewcommand{\arraystretch}{1.3}
\caption{Comparison of area and power estimates of the proposed and contemporary multipliers with $8$ bit precision.}
\resizebox{0.5\textwidth}{!}{
\begin{tabular}{lccccc}
\Xhline{3\arrayrulewidth}
\textbf{Multiplier Type} & \textbf{Latches} & \textbf{Nodes} & \textbf{Edges} & \textbf{Area} & \textbf{Power} \\ \hline \hline
Serial-Parallel \cite{bewick1994fast} & 53 & 274 & 526 & 287.57 & 2808.3 \\
Array \cite{baugh1973two} & 32 & 408 & 780 & 484.59 & 3203.9 \\
Online \cite{ercegovac2004digital} & 62 & 315 & 598 & 313.65 & 3332.5 \\
Online (Pipelined) & 432 & 2385 & 4474 & 2629.39 & 25812.8 \\
Proposed & 315 & 1786 & 3395 & 1947.91 & 18695.5 \\ \Xhline{3\arrayrulewidth}
\end{tabular}}
\label{Comparison2}
\vspace{-3mm}
\end{table}
 
\begin{table}[!ht]
\renewcommand{\arraystretch}{1.3}
\caption{Number of clock cycles required to process $k$ = $8$ vectors of lengths $n$.}
\begin{center}
\vspace{-2mm}
\resizebox{0.5\textwidth}{!}{
\begin{tabular}{lccccc}
\Xhline{3\arrayrulewidth}
\multirow{2}{*}{\textbf{Multiplier Type}} & \multirow{2}{*}{\textbf{Clock Cycles}} & \multicolumn{4}{c}{\textbf{$n$}} \\ \cline{3-6}
 &  & \textbf{8} & \textbf{16} & \textbf{24} & \textbf{32} \\ \hline \hline
Serial-Parallel \cite{bewick1994fast} & $(n+1)*k$& 72 & 136 & 200 & 264 \\
Array \cite{baugh1973two} & $n*k$& 64 & 128 & 192 & 256 \\
Online \cite{ercegovac2004digital}* & $(n+\delta+1)*k$& 96 & 160 & 224 & 288 \\
Online (Pipelined)* & $(n+\delta+1)+(k-1)$ & 19 & 27 & 35 & 43\\ 
Proposed* & $(n+\delta+1)+(k-1)$ & 19 & 27 & 35 & 43\\ 
\Xhline{3\arrayrulewidth}
\multicolumn{4}{l}{\footnotesize *Assuming $\delta = 3$.} \\
\end{tabular}}
\end{center}	
\label{Comparison3}
\vspace{-3mm}
\end{table}

\section{Conclusion}\label{sec: Conclusion}
The proposed method of pipelining allows to remove the unused circuitry which is useful for mitigating static and leakage power and reduction of upto $38$\% and $44$\% in power and area respectively is achieved compared to the pipeline design without truncation of working precision. The area and power savings follow an increasing trend suggesting the suitability of proposed method for larger bit precision. Pipelining of the online multiplier leads to higher throughput and lower latency for streams of input which is highly desirable in deep learning architectures. To compute $8$ input streams of $32$ bits, the serial-parallel, array and non-pipelined online multiplier design requires more than $84\%$, $83\%$ and $85\%$ clock cycles respectively compared to the proposed pipelined design. 

\section{Acknowledgement}
This research was supported by Basic Science Research Program through the National Research Foundation of Korea(NRF) funded by the Ministry of Education(NRF-2020R1I1A3063857) and in part by Korea Institute of Energy Technology Evaluation and Planning, and the Ministry of Trade, Industry and Energy of the Republic of Korea (No. 20184010201650). We also acknowledge the "HPC Support" Project, supported by the ‘Ministry of Science and ICT’ and NIPA in Korea. The EDA tool was supported by the IC Design Education Center (IDEC), Korea.

	\bibliographystyle{IEEEtran}
	\bibliography{Reference}

\end{document}